\documentclass[12pt]{JHEP3}
\usepackage{bm}
\usepackage{amssymb,amsmath,amsthm}
\usepackage{mathrsfs} %script fonts 
\RequirePackage{graphicx}

\newcommand{\be}{\begin{equation}} 
\newcommand{\ee}{\end{equation}}
\newcommand{\bea}{\begin{eqnarray}}
\newcommand{\eea}{\end{eqnarray}}

\newcommand{\gapp}{\mathrel{\raise.3ex\hbox{$>$}\mkern-14mu
              \lower0.6ex\hbox{$\sim$}}}
\newcommand{\lapp}{\mathrel{\raise.3ex\hbox{$<$}\mkern-14mu
              \lower0.6ex\hbox{$\sim$}}}

\newcommand\lsim{\lesssim}
\newcommand\gsim{\gtrsim}

\newcommand\vev[1]{{\langle {#1} \rangle}}
\renewcommand\({\left(}
\renewcommand\){\right)}
\renewcommand\[{\left[}
\renewcommand\]{\right]}

\newcommand\eq[1]{Eq.~(\ref{#1})}
\newcommand\eqs[2]{Eqs.~(\ref{#1}) and (\ref{#2})}

\newcommand\eqsss[4]{Eqs.~(\ref{#1}), (\ref{#2}), (\ref{#3})
and (\ref{#4})}

\newcommand\eqst[2]{Eqs.~(\ref{#1})--(\ref{#2})}

\newcommand\eqreff[1]{(\ref{#1})}
\newcommand\eqsref[2]{(\ref{#1}) and (\ref{#2})}

\newcommand\pa{\partial}

\newcommand\mpl{M_{\rm P}}

\newcommand{\dlabel}[1]{\label{#1}}

\def\calp{{\cal P}}

\def\calpz{{\calp_\zeta}}

\newcommand\bfa{{\mathbf A}}

\newcommand\bfk{{\mathbf k}}

\newcommand\bfw{{\mathbf W}}
\newcommand\bfx{{\mathbf x}}

\newcommand\GeV{\,\mbox{GeV}}

\newcommand\sub[1]{_{\rm #1}}
\newcommand\su[1]{^{\rm #1}}

\newcommand\mone{^{-1}}
\newcommand\mtwo{^{-2}}
\newcommand\mthree{^{-3}}

\newcommand\mfive{^{-5}}

\newcommand\half{^{1/2}}

\newcommand\mn{{\mu\nu}}

\newcommand{\fnl}{f\sub{NL}}

\newcommand\bfkp{{{\bfk}'}}

\newcommand{\phiw}{\phi\sub w}
\newcommand{\rhow}{\rho\sub w}
\newcommand{\tw}{t\sub w}

\newcommand{\calpzphi}{\calp_{\zeta_\phi}}

\newcommand{\zetaphi}{\zeta_\phi}

\newcommand{\zetaw}{\zeta_W}
\newcommand{\hatzetaw}{\hat\zeta_W}

\title {Modulation of the waterfall by a gauge field\footnote
{A preliminary version of this paper appeared as arXiv:1204.6619.}}

\author{David H. Lyth \\
Consortium for Fundamental Physics, Cosmology and Astroparticle Group, 
Department of Physics, Lancaster University, 
Lancaster LA1 4YB, UK \\ 
 E-mail: \email{d.lyth@lancaster.ac.uk}}
\author{Mindaugas Kar\v{c}iauskas \\ 
CAFPE and Departamento de F\'isica Te\'orica y del Cosmos, Universidad
de Granada, Granada-18071, Spain \\ 
E-mail: \email{mindaugas@ugr.es}}

\abstract{We present the first complete calculation of the curvature
perturbation generated during the hybrid inflation waterfall, caused by the 
coupling of the waterfall field to a 
  gauge field $\bfa$ whose kinetic
function $f^2$ depends on the inflaton field. We impose an upper bound on
the  field $\bfw\equiv f\bfa$ which ensures that it has a negligible effect
before the waterfall. We confirm the claim of Soda and Yokoyama, that
the perturbation $\delta \bfw$ generates a statistically
anisotropic spectrum and bispectrum, which could easily be observable.
We also discover a new  phenomenon, whereby the time-dependent
`varyon' field $\bfw$ causes the inflaton contribution to vary during the 
waterfall. The varyon mechanism might be implemented also with a 
scalar field and might  not involve the waterfall.}

\keywords{Primordial curvature perturbation}
\preprint{}

\begin{document}

\section{Introduction}

During the waterfall of hybrid inflation, the perturbation of the waterfall field
generates a contribution  
to the curvature perturbation. But its spectrum
 is proportional to $k^3$ which almost certainly makes it
negligible on cosmological scales \cite{p11}. For the waterfall to 
 generate a  contribution with a nearly flat spectrum, its onset should be modulated 
by (i.e. depend upon the value of) some field 
that is  different from both the inflaton and 
the waterfall field, and whose perturbation has a nearly flat spectrum.

Such  modulation was first  considered in \cite{endinf} using a scalar field
(and further explored in \cite{endapps1,endapps2,endapps3,endapps4}). Then the
Soda and Yokoyama \cite{jiro} used  instead a $U(1)$ gauge field.\footnote
{This is extended to the non-Abelian case in \cite{mind}.}
In this paper we give the first  complete treatment of that case.
The gauge field  $\bfa$ 
has a kinetic function $f^2$ that depends on the inflaton field.
We impose a condition ensuring that $\bfw\equiv f\bfa$  has a 
negligible effect before the waterfall, and take into account both the perturbation of
$f$ and the possible time-dependence of $\bfw$.
We confirm the claim of \cite{jiro}  that the {\em perturbation} $\delta \bfw$
 can generate  a statistical anisotropic contribution to $\zeta$, at a level which 
could easily be observed. We also find a new effect, which is that the 
{\em time-dependence} of $\bfw$ can cause a significant variation in 
 the (statistically isotropic) inflaton contribution  to $\zeta$.
 We dub this new
effect {`the varyon mechanism'} and note that the varyon field  might
 not be a gauge
field and might not act during the waterfall.

We will  take for granted the main ideas of modern cosmology described for
instance in \cite{book}, and use  the  notation and definitions of \cite{ours,book}.
The unperturbed universe has the line element
\be
ds^2 = -dt^2 + a^2(t) \delta_{ij} dx^i dx^j.
\dlabel{ds2}
\ee
In the perturbed universe, we can
 choose a slicing
(fixed $t$) and threading (fixed $\bfx$),  and write for a given quantity
$g(\bfx,t) = g(t) + \delta g(\bfx,t)$.
A different   slicing with a time displacement
$\delta t(\bfx,t)$ gives a different perturbation $\widetilde{\delta g}$.
If $g$ is rotationally invariant we have to first order
\be
\widetilde {\delta g}(\bfx,t) - \delta g(\bfx,t) = 
-\dot g(t) \delta t(\bfx,t)
. \ee
We will invoke this `gauge transformation'  without comment.
In most cases  $g$ is homogeneous on one of the slicings.

 We denote the Fourier component
by $\delta g_\bfk(t)$ where $\bfk$ is the coordinate wavenumber.
Cosmological scales (probed directly by the CMB anisotropy and galaxy surveys)
range from $k=k_0\equiv (aH)_0$ to $k\sim e^{15}k_0$, 
where $H\equiv \dot a/a$
and $(aH)_0$ is evaluated at the present epoch so that $a_0/k_0$ is about the size of the 
observable universe.  A scale is `outside the horizon'
if $k<aH$. 
Inflation corresponds 
to $\epsilon_H< 1$ where $\epsilon_H \equiv -\dot H/ H^2$.
Cosmological scales leave the horizon during inflation and enter the
horizon during the radiation-dominated era leading to Big Bang Nucleosynthesis (BBN).

\section{The curvature perturbation $\zeta$}

\subsection{Definition and $\delta N$ formula}

To define $\zeta$ one smoothes the metric on a super-horizon scale,
and adopts the 
comoving threading and the slicing of uniform energy density $\rho$. Then
\cite{deltan,lms}
\be
\zeta(\bfx,t)\equiv \delta [\ln a(\bfx,t)]
=\delta [\ln \left( a(\bfx,t)/a(t) \right)] \equiv \delta N(\bfx,t)
, \dlabel{deln} \ee
where $a(\bfx,t)$ is the locally defined scale factor (such that a comoving volume element
is proportional to $a^3(\bfx,t)$). The number of $e$-folds
of expansion $N(\bfx,t,t_*)$ starts from a slice at time $t_*$
on which $a$ is unperturbed (`flat slice') and ends
on a uniform $\rho$  slice at time $t$. Since the expansion between two flat slices is uniform,
$\delta N$ is independent of $t_*$.

By virtue of the smoothing, the energy conservation equation is valid locally:
\be
\dot\rho(t) = -3 \frac{\pa a(\bfx,t)}{\pa t} \( \rho(t) + P(\bfx,t) \)
. \ee
In consequence, $\dot\zeta=0$ during an era when $P(\rho)$ is a unique function.
The success of the BBN calculation shows that $P=\rho/3$ to high accuracy 
 just before 
cosmological scales start to enter the horizon. Then  $\zeta$ has a time-independent
value $\zeta(\bfx)$ that 
is strongly constrained by observation. Within observational errors it is gaussian and
statistically isotropic. Its spectrum is nearly
independent of $k$, with \cite{komatsu} 
\bea
\calpz(k) \simeq   (5\times 10\mfive)^2   \dlabel{calpzobs} \\
n(k)-1 \equiv d\ln\calpz/d\ln k = \simeq -0.032\pm 0.012
. \eea
(The result for $n(k)$ assumes that it has negligible scale dependence. It also 
 assumes a tensor fraction $r\ll 10\mone$, which will soon be tested by PLANCK \cite{planck}.)
For the reduced bispectrum \cite{komatsu}
$\fnl$, current observation give $|\fnl|\lsim 100$ and barring a 
detection PLANCK will give $|\fnl|\lsim 10$.  For $\fnl$  to ever be observable we need
$|\fnl|\gsim 1$. 

We will work to first order in $\zeta$, so that
\be
\zeta(\bfx,t) = H(t) \delta t_{f\rho}(\bfx,t)
, \dlabel{delt} \ee
where $\delta t_{f\rho}$ is the time displacement from the flat slice to the 
the  uniform-$\rho$ slice. 
A  second-order calculation of  $\zeta$ is   needed only to treat very small
non-gaussianity corresponding to $|\fnl|\lsim 1$.

We adopt the usual assumption, whereby 
$N(\bfx,t,t_*)$ is determined by the values of one or more fields $\phi_i(\bfx,t)$,
evaluated during inflation at an epoch $t_*$:\footnote
{To be more precise, $N$ will depend also on some of the 
 masses and couplings in the action, and it may depend too 
on the values of any fields with negligible dependence on
$\bfx$ and $t$ that have not time to reach their vacuum expectation values. That does
not affect any of the following.}
\be
N(\bfx,t) = N(\phi_1^*(\bfx),\phi_2^*(\bfx),\cdots,t)
. \ee
Defining the perturbations $\delta\phi_i^*$ on a flat slice, one writes \cite{deltan,lr}
\be
\zeta(\bfx,t) = \sum N_i(t)  \delta\phi_i^*(\bfx) + \frac12 \sum_{ij} 
N_{ij}(t)
\delta\phi_i^*(\bfx) \delta\phi_j^*(\bfx) + \cdots,
\dlabel{deltan} \ee
where  a subscript $i$ denotes $\pa/\pa\phi_i^*$ evaluated at the unperturbed 
point of field space.
The $\phi_i$ are usually taken to be scalar fields, but it has been proposed
\cite{jiro,ours} that some or all of them may be components of a vector field.

On each scale $k$, the 
field perturbations are generated from the vacuum fluctuation at horizon exit
and are initially uncorrelated.
Ignoring scales leaving the horizon after $t_*$  \eq{deltan} defines
a classical quantity $\zeta$.
In general it depends on $t$, settling down to the observed quantity $\zeta(\bfx)$  by
some time $t\sub f$. Since $\zeta(\bfx)$ is nearly gaussian, one assumes that
 \eq{deltan} is  dominated by one or more linear terms involving
nearly gaussian scalar fields.  
With  $t_*$ chosen as the epoch of horizon
exit for a scale $k$ this gives {
\be
\calpz(k,t) \simeq \sum N_i^2(t_*(k),t) \calp_{\delta\phi_i^*(k)}(k,t_*(k))+\ldots 
, \dlabel{deltan2}\ee
}where the terms exhibited correspond to scalar fields, and the dots indicate
vector field contributions \cite{ours}. Each contribution  is positive.

\subsection{Slow-roll inflation}

\dlabel{2.2} 

Slow-roll inflation invokes Einstein gravity, 
and one or more scalar fields with the
canonical kinetic term. The fields have practically gaussian perturbations,
with $\calp_{\delta\phi_i*} = (H/2\pi)^2$ at horizon exit. 
During  single-field slow-roll inflation,  only the inflaton $\phi$ has
significant variation.
Its unperturbed value $\phi(t)$  satisfies 
\be
3H\dot \phi  \simeq -V'(\phi), \dlabel{phidot} 
, \ee
where the potential $V$ satisfies
\bea
\epsilon &\equiv&  \frac12\mpl^2 (V'/V)^2\simeq \epsilon_H \ll 1 \dlabel{eps}\\
|\eta| &\ll 1&,\qquad \eta\equiv \mpl^2 V''/V , \dlabel{eta},
\eea 
giving 
$\rho =  3\mpl^2 H^2 \simeq V$.

The perturbation $\delta\phi_*$ generates a contribution $\zeta_\phi$.
Since $\phi(\bfx,t)$ is the only time-dependent field, the  effect on $N$ of its 
 perturbation $\delta \phi_*(\bfx)$ can be removed 
by the  time shift $\delta t(\bfx)$ which makes $\phi_*$ 
homogeneous, which means that $\zeta_\phi$ is time-independent.
At first order,
\be
\zetaphi(\bfx) = - (H/\dot\phi) \delta\phi_*(\bfx)
, \dlabel{zetaphi} \ee
and 
\bea
\calpzphi(k) &\simeq & \frac1{2\epsilon \mpl^2} \( \frac H{2\pi} \)^2
  \dlabel{calpzphi} \\
n_\phi(k) -1 &\equiv&  d\calpzphi/d\ln k = 2\eta - 6 \epsilon
\simeq 2\eta  \dlabel{npred}
\eea
where the 
 right hand sides are  evaluated at  horizon exit.
The second equality of \eq{npred} is appropriate for small-field models 
\cite{mybound} and it applies
to the standard hybrid inflation which we are going to consider.
The contribution of  $\zeta_\phi$ to $|\fnl|$ is  \cite{maldacena}
$\lsim 10\mtwo$.

For multi-field slow-roll inflation, where two or more fields have significant variation
during inflation, \eqst{zetaphi}{npred}
refer to the contribution of the field pointing
along the inflaton trajectory at horizon exit.
Field perturbations  orthogonal to the (single- or multi-field) 
trajectory give no contribution to $\zeta$ at horizon exit, but may contribute
later. (This may occur during slow-roll inflation in a multi-field model, 
or during the waterfall, or after inflation through a curvaton-type mechanism.)
 We therefore have
\be
\calpzphi(k) \lsim \calpz(k) 
\dlabel{zbound}
,\ee
where $\calpz(k) \simeq (5\times 10\mfive)^2$ is the observed quantity.
The tensor fraction therefore satisfies 
\be
r\leq 16\epsilon
\dlabel{rbound} ,\ee
 with $\epsilon$ evaluated when 
$k_0$ leaves the horizon.\footnote
{This follows from the definition 
$r \equiv \calp_h(k)/\calpz(k)$ with $k\simeq k_0$, and the prediction
$\calp_h(k)=(8/\mpl^2)(H/2\pi)^2$ with $H$ evaluated at horizon exit.}
This leads \cite{mybound} to 
what has been called the Lyth bound, on the variation
$\Delta \phi$ of the inflaton field after $k_0$ leaves the horizon,
$  r \lsim 10\mone   \(\Delta \phi\ / \mpl\)^2 $.
For the tensor fraction to be detectable in the foreseeable future one needs
$r\gsim 10\mthree$ \cite{rlimit}, which is impossible in a small-field model
($\Delta\phi\lsim 10\mone \mpl$).

\section{The model}

\subsection{Hybrid inflation}

\dlabel{earlier}

We are interested only in the era starting with  horizon exit for $k_0$ and ending with
the onset of the 
waterfall. The relevant part of the action is taken to be 
\bea
&&S = \int d^4x \sqrt{-g} \[ \frac12 \mpl^2 R
- \frac12 \pa_\mu \phi \pa^\mu \phi - \frac12 \pa_\mu \chi \pa^\mu \chi 
 - \frac14 f^2(\phi) F_\mn F^\mn  - V \]
, \dlabel{action} \\
&&V(\phi,\chi,A)  =  V_0 + \Delta V(\phi) 
+\frac12 m^2(\phi,A)  \chi^2 + \frac14\lambda \chi^4  
 \dlabel{fullpot2}  \\
&&m^2(\phi,A) \equiv   h^2  A^2 + g^2\phi^2 - m^2
. \dlabel{msq} \eea
with $F_\mn \equiv \pa_\mu B_\nu - \pa_\nu B_\mu$ and  
 $B_\mu$  a $U(1)$ gauge field.
To fix the normalization
of $f$, we set $f=1$ at a time $\tw$ just before the waterfall begins.

Following \cite{ours} we 
 use  the gauge with $B_0=\pa_i B_i=0$, and work with $A_i\equiv
B_i/a$ which is the field defined with respect to the locally orthonormal basis
(as opposed to $B_i$ which is defined with respect to the coordinate basis).
The raised component is $A^i=A_i$ (as opposed to  $B^i=B_i/a^2$).
We also define the canonically normalized field $\bfw\equiv f\bfa$.
The  waterfall field $\chi$ is supposed to be the radial part of a complex field
which is charged under the $U(1)$, generating the first term of \eq{msq}.

 We are going to impose \eq{rhowcon}, which ensures that $\bfw$ has a 
negligible effect before the waterfall. Then, assuming  suitable values for the 
parameters and field values,  \eq{action} gives
what has been called \cite{p11} standard hybrid inflation 
  \cite{andreihybrid,ourhybrid}. 
At each location, the  waterfall begins when
 $m^2(\phi,A)$ falls to zero.   Before it begins,
the waterfall field
$\chi$ vanishes up to a vacuum fluctuation which is set to zero,
 and we have slow-roll inflation with
\be
V = V_0 + \Delta V(\phi)\simeq V_0 
. \ee
 We will take $H$ to be constant which
is typically a good approximation. In contrast with \cite{jiro}, we will not assume
$\Delta V(\phi)\propto \phi^2$. 

During the waterfall, $\chi$ moves to it's vev and then inflation ends.
We will assume that the duration of the waterfall is so short that it can
be taken to occur on a practically unique slice of spacetime. This requires
 \cite{p11} $m\gg H$ and $H\lsim 10^9\GeV$. 

\subsection{Field equations with $f\propto a^\alpha$}

To work out the field equations, {most} previous authors have  taken $f(\phi(\bfx,t))$
to be a function only of time with $f \propto a^\alpha(t)$ {(see however \cite{fprtb})}. 
 Taking  spacetime to be unperturbed, the 
action \eqreff{action} then gives for the unperturbed fields
\bea
\ddot \phi(t) + 3H\dot\phi(t) + V'(\phi(t)) &=& 0 \dlabel{phieq} \\
\ddot \bfw(t) + 3H\dot\bfw(t) + \mu^2 \bfw(t) &=& 0, \dlabel{bfweq}
\eea
where
\be
\mu^2 \equiv H^2 (2+\alpha)(1-\alpha)  \dlabel{mu}
. \ee

By virtue of the flatness conditions \eqsref{eps}{eta},
the first expression is expected to give the slow-roll
approximation \eqreff{phidot} more or less independently of the initial
condition. Similarly, the second equation is expected to give the 
slow-roll approximation { $3H\dot\bfw \simeq - \mu^2 \bfw$ 
if $|\mu|^2 \ll H^2 $. 
This condition is assumed  because the analysis would
otherwise become much more complicated. It is equivalent to
 $\alpha\simeq 1$ or $-2$.

The  first order perturbations satisfy
\bea
\delta \ddot \phi_\bfk(t) + 3H \delta \dot \phi_\bfk(t) + 
\[ (k/a)^2 + V''(\phi(t) \]\delta\phi_\bfk(t) = 0 \dlabel{ddotdelphi}  \\
\delta \ddot \bfw_\bfk(t) + 3H \delta \dot \bfw_\bfk(t) + 
((k/a)^2 + \mu^2)\delta\bfw_\bfk(t) = 0 \dlabel{ddotdelw}
. \eea
Keeping only  super-horizon scales, \eqs{bfweq}{ddotdelw} and give
\be
3H \dot \bfw(\bfx,t) \simeq - \mu^2 delta \bfw(\bfx,t) 
 \dlabel{bfaeq2} .\ee
}The  effect of the metric perturbation (back-reaction) on 
these equations vanishes in the limit where $\phi$ and $\bfw$ are
constant  \cite{book,book1}. The assumption of unperturbed spacetime
is therefore expected to be a good approximation.

In terms of $W$, the coupling $h^2 A^2\chi^2$ becomes
$\tilde h^2 W^2 \chi^2$, where $\tilde h \equiv h/f$. We are 
 setting  $f=1$ when the waterfall begins at $t=\tw$. 
To generate $\delta \bfw$ from the vacuum fluctuation, one  assumes
  that $\bfw$ is  a practically free field while cosmological scales leave
the horizon, corresponding to
$\tilde h\ll 1$,  or  $h\ll e^{-N(k)\alpha}$ where $N(k)$ is the number of $e$-folds
of inflation after horizon exit.
With $\alpha\simeq 1$  this would make
$ h$ too small to have a significant effect.
One therefore assumes  $\alpha \simeq -2$.

The simplest supersymmetric hybrid inflation 
model \cite{qaisar} has $\Delta V$ increasing logarithmically.  Then
$f(\phi)$ increases exponentially. The same behaviour holds for the non-hybrid
model with the full potential $V\simeq 3\mpl^2 H^2 \propto \phi^2$.
It might be  reasonable in string theory
\cite{exponent}, and which could correspond to an  attractor 
\cite{attract}. 

\subsection{Field equations with $f(\phi)$}

In this paper we recognise that $f$ is supposed to be a function of the
inflaton field $\phi$, while retaining  the
assumption $f\propto a^\alpha$ for the unperturbed quantity. 
Using the slow-roll approximation with $\alpha=-2$ we have 
\be
\delta f=\frac {df}{d\phi}\delta\phi = \frac{df}{da} \frac {da}{dt} \frac {dt}{d\phi}
\delta\phi
 = \frac 2 { \sqrt{2\epsilon}  \mpl }\delta\phi
 \dlabel{deltaf} . \ee

Since $f$ is a function of $\phi$, the term $-\frac14 f^2 
F_\mn F^\mn$ in the action couples 
$\phi$ and $\bfw$ so that the right hand sides of 
\eqsss{phieq}{bfweq}{ddotdelphi}{ddotdelw}  are non-zero.
We calculate them in the Appendix, and show that they are negligible
if 
\be
\frac{\rho_W}{\epsilon \rho} =\frac12 \frac{\dot W^2}{\epsilon \rho}
\simeq
\frac12\frac{\mu^2W^2}{\epsilon V} \simeq \frac16 
\frac{ W^2 }{\epsilon \mpl^2}
\ll 1
, \dlabel{rhowcon} \ee
where $\rho_W$ is the energy density of $\bfw$. 
We will assume this condition. It implies
$\rho_W \ll \rho$, which also ensures that $\bfw$ has a negligible effect
during slow-roll inflation.
From \eqs{calpzobs}{calpzphi}  the condition corresponds to
\be
\frac {W(t)}H \lsim 10^5 \( \frac{\calpz(k)}{\calpzphi(k)} \)\half 
, \ee
where $k$ is the scale leaving the horizon at time $t$.

\subsection{The perturbation $\delta \bfw$}

The evolution equation for $\bfw(\bfx,t)$ is the same as that of a 
free scalar field with mass-squared $\mu^2$, and we are assuming
$|\mu|^2 \ll H^2$.
Treating  the Fourier component 
$\delta W_\bfk(t)$ as  an operator and assuming the vacuum state
well before horizon exit, one finds well after horizon exit 
the approximately scale-independent vacuum expectation value
\be
\frac{k^3}{2\pi^2} \vev{ \delta W_\bfk^i(t) \delta W_\bfkp^j(t) }
= \(\delta^{ij} - \hat k^i \hat k^j \)
\delta^3(\bfk+\bfkp) \( \frac H{2\pi } \)^2
  \( \frac k {a(t)  H }  \)^{\frac{2\mu^2}{3H^2}} \dlabel{aspec2}
, \ee
where hats denote unit vectors. From \eq{bfaeq2}, the 
operator  $\delta \bfw_\bfk$ has almost 
constant phase which means that $\delta \bfw_\bfk$
 can be treated as a classical quantity with this correlator.

The decomposition
\be
\bfw(\bfx,t) = \bfw(t) + \delta \bfw(\bfx,t)
\dlabel{deltaW}
\ee
is made in some box of coordinate size $L$
around the observable universe, with $\bfw(t)$
the average within the box.
After smoothing on a cosmological scale $k$, the spatial average of
$(\delta W)^2$ (evaluated within a region not many orders of magnitude
bigger than the observable universe) is of order $\ln (kL) (H/2\pi)^2 $. 
We  assume $W(t) \gg H$,
which is reasonable because $W^2(t)$ at a typical position is expected to be
at least of order the mean square of $(\delta W)^2$ evaluated within
a  box with size $M\gg L$ \cite{mybox}. 

Including both the inflaton and $\bfw$ and assuming that cubic and higher terms
are negligible, \eq{deltan} becomes \cite{ours}
\bea 
\zeta(\bfx,t) &=& \zetaphi(\bfx,t) + \zetaw(\bfx,t) 
+ \sum_i \frac12 N_{\phi\phi}(t) [ \delta\phi_*(\bfx) ]^2
+ N_{\phi i} \delta\phi_*(\bfx) \delta W_i^*(\bfx)  \dlabel{deltanours} \\
\zetaphi(\bfx,t) &\equiv&   N_\phi(t) \delta\phi_*(\bfx), \\
\zetaw(\bfx,t) &\equiv&  \sum_i N_i(t) \delta W_i^*(\bfx) + \frac12 \sum_{ij} 
N_{ij}(t) \delta W_i^*(\bfx) \delta W_j^*(\bfx) 
 , \eea
where the subscripts on $N$ denote partial derivatives evaluated on the unperturbed
trajectory. We are assuming  \eq{rhowcon}, which ensures that before the waterfall
$\zetaw$ is negligible while  $\zeta(\bfx,t)$ is close to 
the time-independent quantity given
by \eq{zetaphi}. 

\section{Effect of the waterfall on $\zeta$}

\subsection{End-of-inflation formula}

Let us denote the contribution generated during the waterfall by $\zeta\sub w$.\footnote
{Notice that we are using the subscript w to indicate the waterfall era. The quantity
$\zeta\sub w$ is the total change in $\zeta$ during the waterfall, which as we shall see
can be different from the contribution $\zetaw$ of $\delta\bfw$.}
To evaluate it,  we assume that the waterfall  happens very quickly  
so that it can be regarded as taking place on a single spacetime
slice. Then \cite{p11,endinf}
\be
\zeta\sub w(\bfx)= H \delta t_{\rho\rho}(\bfx) =  
H\[ \frac{ \delta\rhow(\bfx) }{ \dot\rho(\tw) } - 
\frac{ \delta\rhow(\bfx)}{\dot \rho(t_+) }\] \simeq  H \frac
{ \delta\rhow(\bfx) }{ \dot\rho(\tw) } \simeq H \delta t_{\rho\rm w}
. \dlabel{zwofrho} \ee
In this equation, $ t_{\rho\rho}(\bfx)$ 
is the proper time elapsing between a uniform-$\rho$
slice at time $\tw$ just before the waterfall and a uniform-$\rho$ slice at time
$t_+$ just after the waterfall, while $t_{\rho\rm w}$ is the same thing with the
final slice the waterfall slice itself.

This end-of-inflation formula  actually holds if the waterfall slice is replaced by  
any sufficiently brief  transition from inflation to non-inflation.
In \cite{p11} it is invoked for the transition beginning {\em during} the waterfall,
at the epoch when the evolution of $\chi$ becomes non-linear. 
We are here applying  it to the  entire waterfall.
It   was  first given \cite{endinf} with $A$ in \eq{msq} 
replaced by a   scalar field. In \cite{endinf} the slope of the potential
in the $A$ direction was assumed to be negligible corresponding to
single-field hybrid inflation, and the same assumption was made in several
later papers  \cite{endapps1}. The assumption was relaxed in
 \cite{endapps2,endapps3,endapps4}, 
corresponding to what has been  called \cite{endapps3}  
multi-brid inflation. Following \cite{jiro} we are here taking  
$A$ to be the magnitude of a $U(1)$ gauge field. 
One can also  replace $A$ by a non-Abelian gauge field \cite{mind,toni,mind2}.

\subsection{Waterfall contribution: general formula}

\dlabel{3.2}

Instead of calculating $\zeta\sub w$ directly, we calculate
\be
\zeta(\bfx,t_+) = \zeta\sub w(\bfx) + \zetaphi(\bfx)
, \ee
where $\zetaphi(\bfx)$ is given by \eq{zetaphi}.
We do this first without specifying the function $m^2(\phi,A)$
or the nature of $A$. We  define  $\phiw(A)$ by $m^2(\phiw,A) = 0$. 
(If this equation has more than one solution $\phiw(A)$, we choose one of them.)
The  waterfall occurs when  $\phi(\bfx,t)=\phiw(\bfx,t)$.

If $t_{f\rho}(\bfx)$ is the displacement 
from the  flat  slice at $\tw$ to the uniform $\rho$ slice at $t_+$
we have $\zeta(\bfx,t_+) = H \delta t_{f\rho}(\bfx)$. Making the good approximation
 $\delta t_{f\rho}= \delta t_{f\rm  w}$,
where $t_{f\rm  w}$ is the displacement from the flat slice to the waterfall slice,
we have
\bea
\phi(\bfx,\tw+\delta t_{ f\rho}(\bfx)) 
&=&\phi(\tw) + \delta \phi(\bfx,\tw) + \dot\phi(\tw) \delta t_{ f\rho}(\bfx) \\
\phiw(\bfx,\tw+\delta t_{ f\rho}(\bfx)) 
&=&\phiw(\tw) + \delta \phiw(\bfx,\tw) + \dot\phi\sub w(\tw) \delta t_{ f\rho}(\bfx)
, \eea
where $\delta\phi$ and $\delta\phiw$ are  defined on the flat slice.
For the unperturbed values this gives 
$\phi(\tw)=\phiw(\tw)$. 
For the perturbations it gives 
%we have
\be  
\zeta(\bfx,t_+) = H\delta t_{f\rho}(\bfx) = H \frac{\delta \phiw(\bfx,\tw)
-\delta \phi(\bfx,\tw) }{ \dot \phi(\tw) - \dot\phi\sub w(\tw) }
. \dlabel{zeta} \ee
During hybrid inflation  $\dot\phi<0$, and  we need  $\dot\phi(\tw)< \dot\phiw(\tw)$, 
 or the waterfall will never start.
 
Now we invoke \eq{msq}. 
Discounting  the  strong cancellation $m^2 \simeq h^2 A^2$ it gives
\be
~ \phiw(\bfx,t)  = \frac1g (m^2- h^2 A^2 (\bfx,t) )\half 
\simeq  \frac mg - \frac12 \frac{h^2 A^2(\bfx,t) }{mg} 
.  \dlabel{phiw2} \ee

In most previous work, $A$ is taken to be a scalar field. 
For single-field 
hybrid inflation 
\cite{endinf,endapps1},   $\dot\phi\sub w$  is supposed to be negligible.
Then $\phiw$ has a practically time-independent value and the waterfall slice 
corresponds to simply $\phi(\bfx,t)= \phiw(\bfx)$. For two-brid inflation
\cite{endapps2,endapps3,endapps4},   $\phi$ and $A$ have equal status and 
 the time-dependence of $\phiw$ is significant. 
We have checked that in this case,  \eqs{zeta}{phiw2} are equivalent to the result
(4.1) given in \cite{endapps3}.

In our case  $A$ is the magnitude of a $U(1)$ gauge field with the action \eqreff{action}. 
Let us first follow \cite{jiro} by setting $\alpha=-2$.
 {}From \eq{bfaeq2}, this
 makes $W(\bfx,t)$ time-independent. Then, if we 
 ignore the perturbation $\delta f$ we have $A(\bfx,t)\propto 1/f \propto a^2$.
This gives \eq{deltanours} for $\zeta(\bfx,t_+)$, with\footnote
{We have checked that in this and the following cases, the third and fourth terms
of \eq{deltanours} are negligible.} 
\bea
\zetaphi(\bfx,t_+) &=& \frac{ \zetaphi(\bfx)}{1-X},  \\
\zetaw(\bfx,t_+) &=&  \frac {\hatzetaw(\bfx)}{1-X}, \dlabel{z1}
\eea
where 
\be
X\equiv \frac{ h^2 W^2(\tw) }{\mpl m g \sqrt{2\epsilon(\tw)} }
, \ee
and 
\be 
\hatzetaw(\bfx) = 
- (X/W^2(\tw) ) \( \bfw(\tw)  \cdot \delta\bfw(\bfx,\tw) + \frac12
 \delta\bfw(\bfx,\tw) \cdot \delta\bfw(\bfx,\tw) \).
\ee
(We display 
$\tw$ for future reference, even though we are for the moment taking
$W$ to be time-independent.) 
We need $X<1$ for the waterfall to end.

Ignoring the time-dependence of {$\bfa$} we have $\zetaw(\bfx,t_+)
=\hatzetaw(\bfx)$, which is the  result obtained in  \cite{jiro}.
We see that  $\zetaw$ becomes bigger when the time-dependence is taken into account.
The effect of the time-dependence of {$\bfa$} was previously considered in \cite{hassan},
who conclude that it {\em decreases} 
$\zetaw$  by a factor $e^{-2N(k)}$ making it far too small to have 
an observable effect. But the  calculation of \cite{hassan} is not from first principles
because $\bfa$ is treated as a scalar field. 

We have yet to include $\delta f$. This has no effect on 
 $\zetaw$, but for $\zetaphi$ it cancels the effect of $\dot A$ giving
\be
\zetaphi(\bfx,t_+) = \zetaphi(\bfx) \dlabel{z2}
. \ee

In the above we worked with $\phiw(A)$ defined by $m^2(\phiw,A)=0$.
That  allows comparison with previous work
where  $A$ is a scalar field, and with $\cite{jiro,hassan}$ where $A$ is the magnitude
of a gauge field. But in the latter case the calculation becomes simpler if
we use instead $\phiw(W)$ defined by $m^2(\phiw,(W/f(\phiw))=0$, because it is $\bfw$
that decouples from $\phi$. With our current assumption $\mu^2=0$, $\dot \bfw$
vanishes which means that $\dot\phiw(W)=0$. 
Evaluating $\delta \phiw(W)$ we again arrive at \eqs{z1}{z2}.
 According to this derivation, the factor $1-X$ which was absent in \cite{jiro}
 comes from $\delta f$. 

\subsection{The varyon mechanism}

Allowing $\mu^2\neq 0$, we have $\dot W\neq 0$ and hence 
 $\dot\phiw(W)\neq0$. This gives\footnote
{In \cite{p12}, this formula is given with $1-X$ incorrectly replaced by $1+X$.
That makes the first term must be close to 1 whatever the value of $X$.}
\be
\zeta(\bfx,t_+) =  
\[1+ \frac{\mu^2}{6H^2} \frac{X}{1-X} \]\mone
\[ \zetaphi(\bfx)    +  \frac{\hatzetaw(\bfx)}{ 1-X } \]
\dlabel{main} \ee
We see that 
 $\zetaphi(\bfx,t)$ 
 is altered by the waterfall, which seems to contradict 
 the statement in Section \ref{2.2},
that the slow-roll inflation result  $\zetaphi(\bfx)$ given by \eq{zetaphi} persists 
even after slow-roll ends. 
There is in fact no contradiction, because the presence of $\bfw(t)$ means that 
we are not dealing with exact slow-roll inflation. As a result, 
 $\phi(\bfx,t)$ is not the only time-dependent field, and the  effect on 
$N(\bfx,t,t_*)$ of its 
 perturbation $\delta \phi_*$ is not removed 
by the  time shift which makes $\phi_*$ 
homogeneous. 
Note  that the perturbation $\delta\bfw_*(\bfx)$ is in this context irrelevant. It is the 
quantity $\dot W(t)$ that is causing the effect.

In our case $W$ is the magnitude of 
a gauge  field and its effect on $\zeta$ occurs during the waterfall.
But in general, 
any time-dependent field might do the same thing, i.e. have a negligible effect
during slow-roll inflation but a significant one later owing to the time-dependence
of its unperturbed part. The only exception is if the field
is a  slowly rolling scalar field; in that case its effect is just to slightly alter 
the direction in field space of the trajectory, which as we discussed in Section
\ref{2.2} will still give the (almost unchanged)
the time-independent perturbation $\zetaphi(\bfx)$. 
We propose to call a field causing the new effect a {\em varyon}.

The varyon mechanism can remove the bound \eqreff{zbound} and hence the bound
\eqreff{rbound} on the tensor fraction. That might allow
an observable tensor fraction within a small-field inflation model.
The possibility of avoiding \eq{rbound}  was mooted in \cite{bkr,martin}
but they did not find a mechanism. One can easily avoid \eq{rbound}
by abandoning the canonical kinetic term for the inflaton
\cite{slava} and we are now,  for the
first time, pointing
to a possible mechanism with the canonical kinetic term. 

In our case, the varyon $\bfw$ has only a small effect except in the
 the very fine-tuned regime
$(1-X) \ll |\mu^2|/H^2$. Even if its effect is significant, and reduces
 $\zetaphi$ (positive  $\mu^2$) so as to give $r>12\epsilon$,
it cannot make $r$  big enough to observe  because the
end-of-inflation formula requires $H\lsim 10^9\GeV$ corresponding to 
$r<10^{-10}$.
It remains to be seen if a different varyon, perhaps a scalar field, can give a more
interesting result.

\subsection{Anisotropic spectrum and bispectrum}

 Since we are taking  $W \gg  H$,  
the linear  term of $\zetaw$ dominates,
 leading to  a spectrum of the form \cite{jiro,ours}
\be
\calpz(\bfk) = 
\calp_\zeta(k)  \[ 1 -  \beta \( \hat \bfa \cdot \hat \bfk \)^2 \] \\
.  \ee
On cosmological scales, observation  requires $|\beta|\lsim 10\mone$, 
and barring a detection PLANCK
will give $|\beta| \lsim 10\mtwo$ \cite{anis}. We therefore have
$\calpz(\bfk)\simeq \calpz(k)$ with (\eq{calpzobs})
$\calpz(k)\simeq (5\times 10\mfive)^2$.

With the parameters constrained to give $\beta\ll 1$, we have
\cite{jiro,ours} 
\bea
\beta &=& \frac{h^4 W^2(\tw)}{m^2 g^2} \frac{\epsilon(t_k)} {\epsilon(\tw) }
\nonumber \\
&\times& \frac{\calpzphi(k) }{\calpz(k) } \( 1- X \)\mtwo \nonumber \\
&\times & e^{-\frac{2\mu^2}{3H^2} N(k) } 
\[1+ \frac{\mu^2}{6H^2} \frac{X}{1-X} \]\mtwo
, \dlabel{betapred} \eea
where $t_k$ is the epoch of horizon exit for the scale $k$, and
$\calpzphi$ is given by \eq{calpzphi}.
The first line is the result of \cite{jiro}, who assumed $\calpz(k)=\calpzphi(k)$.
The first factor of the second line drops that assumption while the second factor
takes account of the  inhomogeneity of $f(\phi)$.
The third line allows $\mu\neq 0$.

Including the quadratic term of  $\hat \zeta\sub w$ we reproduce the result of
\cite{jiro} for the  reduced bispectrum:
\be
\fnl = f\sub{NL}\su{iso} \( 1 + 
f\su{ani}(\bfk_1,\bfk_2,\bfk_3) \)
\ee
where   
\bea
f\su{ani} &=& \frac{
 - (\hat \bfa \cdot \hat \bfk_1)^2
-(\hat \bfa \cdot \hat \bfk_2)^2 + (\hat \bfk_1 \cdot \hat \bfk_2)
(\hat \bfa \cdot \hat \bfk_1)(\hat \bfa \cdot \hat \bfk_2)  }{
\sum k_i^3/ k_3^3  }
+ \mbox{ 2 perms.} \\
 f\sub{NL}\su{iso} &=&   \frac53 \frac{\beta^2}{  X } 
. \dlabel{fnl} \eea

\section{Conclusion}

Although there 
is so far no evidence for 
statistical  anisotropy of the primordial curvature perturbation $\zeta$, 
mechanisms have been proposed   for generating it.
 Most of them  invoke a  vector field.

One mechanism takes  the vector field to be  homogeneous during inflation, 
but causes significant anisotropy in the  expansion \cite{jiro1} 
(for a recent review of this approach see \cite{jirorev}).  
Then the perturbations of scalar fields generated from the vacuum fluctuation will be 
statistically anisotropic, and so too will be $\zeta$  on the usual assumption that 
it originates from one or more of these perturbations.

A different mechanism 
takes  the inflationary expansion to be  practically isotropic, but generates a
vector field perturbation from the vacuum fluctuation
 \cite{jiro,ours}
(for the most recent paper on this approach see \cite{km}).\footnote
{The use of a vector field to generate a contribution to $\zeta$ was first mooted in
\cite{kostas}.} 

In this paper we have given the first complete treatment of the
 version of the second  mechanism
proposed in \cite{jiro}, which couples the waterfall  field to a gauge field
$\bfa$ whose kinetic function $f^2$ depends on the inflaton.
 We have confirmed their claim 
that the statistical anisotropy could easily
be big enough to observe, and we have also discovered
 a completely new effect; if $\bfw\equiv f\bfa$ is time-dependent it causes the
usual 
{\em inflaton} contribution  $\zeta_\phi$ to vary during the waterfall. This `varyon'
effect might  still occur if $\bfw$ 
is replaced by a time-dependent  (but not slowly rolling)
{\em scalar} field and it might have nothing to do with the waterfall.

%%%%%%%%%%%%%%%%%%%%%%%%%%%%%%%%%%%%%%%%%%%%%%%%%%%%%%%%%%%%%%%%%%%%%%
\section{Acknowledgments}
%%%%%%%%%%%%%%%%%%%%%%%%%%%%%%%%%%%%%%%%%%%%%%%%%%%%%%%%%%%%%%%%%%%%%%
DHL 
 acknowledges support from the Lancaster-Manchester-Sheffield Consortium for
Fundamental Physics under STFC grant ST/J00418/1, and from
 UNILHC23792, European Research and Training Network (RTN) grant. 
MK is supported by the grants CPAN CSD2007-00042
and MICINN (FIS2010-17395).
We thank  K.~Dimopoulos, H.~Firouzjahi, 
J.~Soda and  S.~Yokoyama for discussion in the early stage of this work. 

%%%%%%%%%%%%%%%%%%%%%%%%%%%%%%%%%%%%%%%%%%%%%%%%%%%%%%%%%%%%%%%%%%%%%
\appendix
%%%%%%%%%%%%%%%%%%%%%%%%%%%%%%%%%%%%%%%%%%%%%%%%%%%%%%%%%%%%%%%%%%%%%
\section{Equations of Motion for $\phi(\bfx,t)$ and $\mathbf W(\bfx,t)$}

Extremizing the action in \eq{action} with respect to fields $\phi$, $B_\mu$ and their derivatives we obtain field equations 
\begin{eqnarray}
\left[\partial_{\mu}+\partial_{\mu}\ln\sqrt{-g}\right]\partial^{\mu}\phi+V'+
\frac{1}{2}ff'F_{\mu\nu}F^{\mu\nu}&=&0;\\
\left[\partial_{\mu}+\partial_{\mu}\ln\sqrt{-g}\right]fF^{\mu\nu}&=&0,
\end{eqnarray}
where $g\equiv \mathrm{det} (g_{\mu\nu})$ and 
$f'\equiv\partial f/\partial\phi$. Choosing the temporal gauge $B_0 = 0$ and a line element of the unperturbed universe in \eq{ds2}, one finds equations of motion for the fields $\phi(\bfx,t)$ and $\mathbf B(\bfx,t)$
\begin{eqnarray}
\ddot\phi+3H\dot\phi-a^{-2}\nabla^{2}\phi+V' & = & -\frac{1}{2}f f' F_{\mu\nu}F^{\mu\nu}, \label{EoM-phi} \\
\ddot B_i+\left(H+2\frac{\dot f}{f}\right)\dot{B}_{i}-a^{-2}\nabla^{2}B_{i} & = & a^{-2}2\frac{\partial_{j}f}{f}\partial_{j}B_{i}, \label{EoM-W}
\end{eqnarray}
Recasting the above equations in terms of $\mathbf W \equiv f \mathbf B /a$ and dropping gradient terms, one arrives at equations of motion for homogeneous fields $\phi(t)$ and $\bfw (t)$
\begin{equation}
\ddot\phi+3H\dot\phi+V'=\frac{f'}{f}\left[\dot{\mathbf{W}}+\left(H-\frac{\dot{f}}{f}\right)\mathbf{W}\right]^{2}, \dlabel{hom-phi}
\ee
\be
\ddot{\mathbf{W}}+3H\dot{\mathbf{W}}+\left(2H^{2}-H\frac{\dot{f}}{f}-\frac{\ddot{f}}{f}\right)\mathbf{W}=0, \dlabel{hom-W}
\end{equation}
where we also used $\dot H \simeq 0$.

Decomposing the field $\bfw(\bfx,t)$ as in \eq{deltaW} and similarly the field $\phi(\bfx,t)$,  we find equations of motion for  $\delta\phi(\bfx,t)$ and $\delta \bfw(\bfx,t)$ from \eqs{EoM-phi}{EoM-W}. Keeping only the first order terms and switching to the Fourier space they become
\begin{eqnarray}
& &\delta \ddot\phi_\bfk +3H\delta\dot\phi_\bfk+\left(\frac{k^2}{a^2} +V''\right)\delta\phi_\bfk = \nonumber\\
& & \quad\quad = 2\frac{f'}{f}\left[ \dot\bfw +\left(H-\frac{\dot{f}}{f}\right)\bfw \right] \left[\delta\dot\bfw_\bfk+\left(H-\frac{\dot{f}}{f}\right) \delta \bfw_\bfk - \delta\left(\frac{\dot{f}}{f}\right)_\bfk \bfw\right], \dlabel{prtb-phi}\\
& &\delta\ddot\bfw_\bfk + 3H\delta\dot\bfw_\bfk + \left(2H^{2}-H\frac{\dot{f}}{f}-\frac{\ddot{f}}{f}\right)\delta\bfw_\bfk + \frac{k^2}{a^2} \delta\bfw_\bfk = \nonumber\\
& & \quad\quad = \left[ H\delta\left(\frac{\dot{f}}{f}\right)_\bfk + \delta\left(\frac{\ddot{f}}{f}\right)_\bfk + \frac{f'}{f}\, \frac{k^2}{a^2}\delta\phi_\bfk\right]\bfw.\dlabel{prtb-W}
\end{eqnarray}
With our choice $f\propto a^\alpha$ for the unperturbed $f$, the above expressions become
\begin{eqnarray}
& &\delta\ddot\phi_\bfk + 3H\delta\dot\phi_\bfk + \left(\frac{k^2}{a^2}  + V''\right) \delta\phi_\bfk =\nonumber \\
& &\quad\quad = \frac{-2\alpha}{\sqrt{2\epsilon}\mpl} 
                 \left[\dot\bfw + H \left(1-\alpha\right)\bfw \right] 
                 \left[\delta\dot\bfw_\bfk + H\left(1-\alpha\right) \delta\bfw_\bfk - \alpha\bfw \frac{H}{\dot{\phi}} \delta\dot\phi_\bfk \right],
\dlabel{a9}  \\
& &\delta\ddot\bfw_\bfk + 3H\delta\dot\bfw_\bfk + \left(\frac{k^2}{a^2}+\mu^2 \right) \delta\bfw_\bfk = \nonumber \\
& &\quad\quad =\frac{-\alpha \bfw}{\sqrt{2\epsilon}\mpl}
                \left[ \delta\ddot\phi_\bfk + H \left(1+2\alpha\right) \delta\dot\phi_\bfk +\frac{k^2}{a^2} \delta\phi_\bfk \right], \dlabel{a10}
.
\end{eqnarray}

The energy density of the vector field in \eq{action} is given by \cite{fF2curv} $\rho_{B}(\bfx,t)=-f^{2}F_{\mu\nu}F^{\mu\nu}/4$. From this it is easy to see that the background value of $\rho_{B}(\bfx,t)$ is given by
\be
\rho_{B}(t)=\frac{1}{2}f^{2}\left(\frac{\dot{B}}{a}\right)^{2}=\frac{1}{2}\left[\dot{W}+\left(H-\frac{\dot{f}}{f}\right)W\right]^{2}\simeq\frac12 H^2 W^2.
\ee
The right hand side of 
 Eq.\eqref{hom-phi}
is negligible if $\rho_{B}$ satisfies \eq{rhowcon}. 
We now show that the same is true of the right hand sides of
\eqs{a9}{a10}.
At the epoch $k\sim aH$,  the terms on the left hand sides
are of order $H^3$ and \eq{rhowcon} ensures that the right hand sides
are indeed much smaller. At the epoch $aH/k =\exp(N_k(t))\gg 1$,
the first term of each left hand side is negligible.
The other two terms are of order $|\eta|\equiv |V''|
/3H^2$ for \eq{a9}  and of order $|\eta_W|\equiv |\mu^2|/3H^2$
for \eq{a10}.
\eq{rhowcon} ensures that the right hand side of \eq{a9} is negligible,
and it ensures that the right hand side of \eq{a10} is negligible
if also $|\eta_W|\gg 10\mfive$. But the latter condition is irrelevant, because
its violation makes the time-dependence of $W$ (coming then from
the right hand side) negligible.

%%%%%%%%%%%%%%%%%%%%%%%%%%%%%%%%%%%%%%%%%%%%%%%%%%%%%%%%%%%%%%%%%%%%%%
%\bibliographystyle{unsrt}

\end{document}